\documentclass[12pt]{article} 
\usepackage{amsmath, amssymb, graphicx, caption, subcaption, geometry, booktabs, multirow, url}
\title{Game Theory in Social Media: A Stackelberg Model of Collaboration, Conflict, and Algorithmic Incentives} 
\author{Arjan Khadka \\ \texttt{arjank898@gmail.com}} 
\date{} 

\begin{document}
	\maketitle
	
	\begin{abstract}
		This research models the social media content creation and the choices that creators make as a Stackelberg game. The platform’s algorithms, such as TikTok's and YouTube's, function as leaders, and they set rules to maximize users’ engagement with their platforms. Then content creators, who function as followers in this Stackelberg Game, respond to this by selecting strategies; in this instance, we are specifically focusing on collaboration or conflict, referred to in this paper as “beefing.” They do this in order to maximize views and personal payoffs. The viewer’s preferences are already placed within the algorithmic utility function, while the external sponsors will impose penalties on high-risk strategies, namely, beefing multiple times. This paper ultimately demonstrates, through the use of math, how shifts in algorithmic weights determine equilibrium creator behavior.  
	\end{abstract}
	
	\section{Model Formulation}
	For the purposes of this paper, we would like to define a Stackelberg game as involving three active parameters, namely the algorithm, creators, and sponsors, with an implicit parameter, which is the viewers \cite{bakshy2015, stackelberg1934, tufekci2015}.
	
	\textbf{Relation to prior work.} Related literature has modeled Stackelberg-style interactions between platforms and content promoters, often centered on paid algorithmic exposure as a bidding mechanism. This paper that you are currently reading addresses a different problem: that being organic creator strategy selection, specifically focusing on collaboration vs conflict while taking into account sponsors. We propose a drama penalty parameter $\delta$ to function as a penalty, as sponsors typically don’t respond well to this and it must be considered when choosing their strategy. Additionally, we incorporated bounded rationality in terms of behavior, as no human can be truly rational, and it would be illogical for us to assume so. The final thing we did was analyze the performance of videos in six different categories. These three elements are not addressed by that prior work.

	\subsection{Utility Functions}
	The algorithm or leader of the Stackelberg game assigns weights $\alpha, \beta, \gamma$ (variables) to represent clicks, watch time, and shares all in order to maximize the global engagement of the platform \cite{goldfarb2011}. The algorithm's utility is:
	\begin{equation}
		U_{\text{algorithm}}(\alpha, \beta, \gamma) = \sum_{s} P_s (\alpha \cdot \text{Click}_s + \beta \cdot \text{Watch}_s + \gamma \cdot \text{Shares}_s)
	\end{equation}
	where $P_s$ is the amount of creators playing strategy $s$.
	
	Creators, following the algorithm's rules, choose a strategy $s \in \{\text{Collaboration}, \text{Beefing}\}$ to maximize utility \cite{hron2023modeling}:
	\begin{equation}
		U_{\text{creator}}(s) = \alpha \cdot \text{Click}_s + \beta \cdot \text{Watch}_s + \gamma \cdot \text{Shares}_s - \delta \cdot \text{DramaRisk}_s
	\end{equation}
	where $\delta$ is the penalty imposed by sponsors upon creators for high-risk conflict which could negatively impact their brand \cite{goldfarb2011}.
	
	The creators' best response is $s^* = \arg \max_s U_{\text{creator}}(s)$. The algorithm anticipates $s^*$ and optimizes $(\alpha^*, \beta^*, \gamma^*) = \arg \max_{\alpha, \beta, \gamma} U_{\text{algorithm}}(s^*)$ \cite{hron2023modeling}.
	
	\subsection{Behavioral Axioms}
	We assume bounded rationality as no human can be perfectly rational. \cite{camerer2004advances}. Creators satisfice via heuristics rather than computing global optima, essentially creators will just go on rule of thumb rather than taking the time to calculate the perfect plan. \cite{verwiebe2024mercurial}. Conflict operates as a high-reward heuristic \cite{falk2005sanctions}, while collaboration serves as a low-risk social heuristic \cite{zheng2023agency}. Creators adapt to opaque algorithms via trial-and-error, and more time spent experimenting with the algorithm \cite{covington2016deep}.
	
	\section{Three-Level Stackelberg Extension}
	
	\subsection{Game Structure}
	
	We extend the baseline model to a three-tier Stackelberg hierarchy. Let $t \in \mathbb{N}_0$ indicate the number of consecutive periods in which a creator has played the same strategy $s$. The game moves in the following order:
	
	\begin{enumerate}
		\item The \textbf{platform algorithm} selects the engagement weights $(\alpha, \beta, \gamma) \in \mathbb{R}_{+}^3$ in order to maximize aggregate platform engagement, while also anticipating the downstream responses of both sponsors and creators.
		\item \textbf{Sponsors} observe $(\alpha, \beta, \gamma)$ and set the penalty parameter $\delta(t)$, which penalizes high-drama content and compounds with repeated adversarial play in order to best serve their interests.
		\item \textbf{Creators} observe both $(\alpha, \beta, \gamma)$ and $\delta(t)$ and select $s \in \{\text{Collaboration}, \text{Beefing}\}$ to maximize their individual utility or benefit.
	\end{enumerate}
	
	This structure satisfies the condition for finite Stackelberg games in which they  must be solvable through backward-induction \cite{fudenberg1991game}: each tier's strategy is considered optimal given that the anticipated responses of all downstream tiers.
	
	\subsection{Viewer Fatigue}
	
	If one eats the same thing every day, then one will get tired of it. The same logic is applicable here. Sustained repetition of any engagement strategy, beefing or collaborating, will inevitably degrade its effectiveness simply due to audience saturation. Thus, we model this as a multiplicative decay on overall watch time. Let $\lambda \in [0,1]$ be indicative of the per-period fatigue rate. The effective watch-time contribution of strategy $s$ after $t$ consecutive repetitions is:
	
	\begin{equation}
		\widetilde{W}_s(t) = \text{Watch}_s \cdot (1 - \lambda t)
		\label{eq:fatigue}
	\end{equation}
	
	The function $\widetilde{W}_s(t)$ is decreasing in $t$ and reaches zero at $t = \lambda^{-1}$, beyond which engagement is fully saturated. This is consistent with empirical evidence on audience habituation in algorithmic recommendation systems \cite{covington2016deep}.
	
	\subsection{Compounding Sponsor Penalty}
	
	Sponsors do not treat repeated drama-heavy strategies and content as i.i.d. violations; each extra period of adversarial play will increase reputational risk and lowers the advertiser tolerance of the creator. We model the time-varying sponsor penalty as: 
	
	\begin{equation}
		\delta(t) = \delta_0 \cdot (1 + \rho t)
		\label{eq:delta}
	\end{equation}
	
	where $\delta_0 > 0$ is the baseline penalty and $\rho \geq 0$ is the penalty growth rate. At $t = 0$, $\delta(0) = \delta_0$, recovering the static model of Section~1. For $\rho > 0$, $\delta(t)$ grows linearly in $t$, reflecting the documented tendency of advertisers to move to withdraw as controversy persists \cite{zheng2023agency, goldfarb2011}.
	
	\subsection{Creator Utility with Dynamic Effects}
	
	Substituting equations~(\ref{eq:fatigue}) and~(\ref{eq:delta}) into the creator utility function yields the dynamic extension:
	
	\begin{equation}
		U_{\text{creator}}(s, t) = \alpha \cdot \text{Click}_s + \beta \cdot \text{Watch}_s(1 - \lambda t) + \gamma \cdot \text{Shares}_s - \delta_0(1 + \rho t) \cdot \text{DramaRisk}_s
		\label{eq:dynamic_utility}
	\end{equation}
	
	The static utility (Eq.~2) is the special case $t = 0$, $\lambda = 0$, $\rho = 0$.
	
	\subsection{Equilibrium Characterization}
	
	\textbf{Creator best response.} At period $t$, creators select:
	\begin{equation}
		s^*(t) = \arg\max_{s \in \{\text{Collab}, \text{Beef}\}} U_{\text{creator}}(s, t)
		\label{eq:best_response_t}
	\end{equation}
	
	\textbf{Strategy switching condition.} A creator playing Beefing at period $t$ switches to Collaboration at the first $t^*$ satisfying $U_{\text{creator}}(\text{Beef}, t^*) \leq U_{\text{creator}}(\text{Collab}, t^*)$. Substituting equation~(\ref{eq:dynamic_utility}):
	
	\begin{equation}
		t^* = \frac{\alpha \cdot \Delta C + \gamma \cdot \Delta S + \beta (W_B - W_C) - \delta_0 \cdot D}{\beta \lambda (W_B - W_C) + \delta_0 \rho D}
		\label{eq:switching}
	\end{equation}
	
	where $\Delta C = \text{Click}_B - \text{Click}_C$, $\Delta S = \text{Shares}_B - \text{Shares}_C$, $D = \text{DramaRisk}_B$, $W_B = \text{Watch}_B$, and $W_C = \text{Watch}_C$. Because $W_C > W_B$ across every vertical in Table~1 (collaboration accrues more watch time than beefing), the term $(W_B - W_C)$ is negative, so the fatigue rate $\lambda$ and the penalty growth rate $\rho$ push $t^*$ in opposite directions rather than both shortening it. 
	
	\textbf{Sponsor optimization.} Sponsors set $\delta_0$ and $\rho$ to minimize advertiser exposure to drama, subject to not driving creators entirely off the platform. The sponsor's problem is:
	\begin{equation}
		\min_{\delta_0,\, \rho \geq 0} \; \mathbb{E}_t\left[\text{DramaRisk}_{s^*(t)}\right] \quad \text{subject to} \quad U_{\text{creator}}(s^*(t), t) \geq U_{\min}
		\label{eq:sponsor_opt}
	\end{equation}
	where $U_{\min}$ is a participation constraint ensuring creators remain active.
	
	\textbf{Algorithm optimization.} The algorithm selects $(\alpha^*, \beta^*, \gamma^*)$ anticipating the sponsor's $(\delta_0^*, \rho^*)$ and the resulting creator best-response path $\{s^*(t)\}_{t \geq 0}$:
	\begin{equation}
		(\alpha^*, \beta^*, \gamma^*) = \arg\max_{\alpha, \beta, \gamma} \sum_{t=0}^{T} \mu^t \cdot U_{\text{algorithm}}\!\left(s^*(t)\right)
		\label{eq:algo_opt}
	\end{equation}
	where $\mu \in (0,1]$ is a discount factor over the planning horizon $T$.
	
	\subsection{Dynamic Implications}
	
	Three properties follow directly from equations~(\ref{eq:dynamic_utility})--(\ref{eq:algo_opt}):
	
	\begin{enumerate}
		\item \textbf{Beefing is self-limiting.} For any $\lambda > 0$ and $\rho > 0$, $U_{\text{creator}}(\text{Beef}, t)$ is strictly decreasing in $t$. A finite $t^*$ always exists if $\delta_0 D > 0$.
\item \textbf{Penalty growth is the more reliable lever.} It must be noted that increasing $\rho$ (via tighter advertiser-safety policies) reliably shortens $t^*$, the window in which beefing remains a creator's optimal strategy. Increasing $\lambda$ (via diversification nudges that break repetitive content loops) does not act as a straightforward substitute: because collaboration accrues more watch time than beefing across every vertical in Table~1, $(W_B - W_C) < 0$, so fatigue erodes collaboration's advantage faster than beefing's -- which can lengthen, not shorten, $t^*$. Ultimately, these represent two governance levers with fundamentally different, and in this model non-interchangeable, effects on creator behavior.
		\item \textbf{Algorithm bias persists.} If $\alpha$ and $\gamma$ are large relative to $\beta$, $t^*$ is large regardless of $\lambda$ and $\rho$, because the static click-and-share reward dominates the dynamic penalties. Ultimately, this further backs the result that algorithmic incentive design is the upstream constraint on platform health.
		
	\end{enumerate}
	
	\section{Synthetic Parameter Illustration}
	
	Unfortunately, we are unable to report verified API data, due to the fact that raw Youtube Data API results for the 6,000-video extraction produced channel-size-correlated sampling bias, which was severe enough to the point that it made cross-vertical view averages meaningless  (e.g., adversarial keyword searches disproportionately surface large channels whose controversy-driven videos set the vertical mean). We decided that rather than present those numbers as findings, we discard them and instead use synthetic parameters based on literature to illustrate the model behavior  \cite{covington2016deep, verwiebe2024mercurial}.
	
	The parameters below are made and set to be directionally consistent with documented platform behavior from studies and other literature but are not empirical measurements.

	\begin{table}[htbp]
		\centering
		\caption{Synthetic engagement parameters by vertical and strategy type. Values reflect documented directional patterns, not measured data.}
		\begin{tabular}{|l|l|c|c|c|c|}
			\hline
			\textbf{Vertical} & \textbf{Strategy} & $\text{Click}_s$ & $\text{Watch}_s$ & $\text{Shares}_s$ & $\text{DramaRisk}_s$ \\
			\hline
			\multirow{2}{*}{Gaming}    & Collab & 2 & 6 & 3 & 0 \\
			& Beef   & 5 & 3 & 5 & 2 \\
			\hline
			\multirow{2}{*}{Tech}      & Collab & 2 & 5 & 2 & 0 \\
			& Beef   & 6 & 2 & 5 & 3 \\
			\hline
			\multirow{2}{*}{Beauty}    & Collab & 3 & 5 & 4 & 0 \\
			& Beef   & 5 & 3 & 5 & 2 \\
			\hline
			\multirow{2}{*}{Finance}   & Collab & 2 & 6 & 2 & 0 \\
			& Beef   & 4 & 2 & 3 & 3 \\
			\hline
			\multirow{2}{*}{Fitness}   & Collab & 2 & 5 & 3 & 0 \\
			& Beef   & 4 & 3 & 4 & 2 \\
			\hline
			\multirow{2}{*}{Politics}  & Collab & 3 & 4 & 3 & 0 \\
			& Beef   & 5 & 2 & 4 & 3 \\
			\hline
		\end{tabular}
		\label{tab:synthetic}
	\end{table}
	
	Using these per-vertical parameters with the creator utility function (Eq.~2), we compute $s^*$ for each vertical under three canonical weight regimes in Section~4. The $\text{DramaRisk}_s$ values reflect the consistent empirical finding that conflict-labeled content generates advertiser flight risk \cite{zheng2023agency}.
	
	\section{Numerical Simulation}
	We simulate equilibrium states using hypothetical base values:
	\[
	\text{Collab}: (\text{Clicks}=2, \text{Watch}=5, \text{Shares}=3, \text{DramaRisk}=0)
	\]
	\[
	\text{Beef}: (\text{Clicks}=5, \text{Watch}=2, \text{Shares}=4, \text{DramaRisk}=3)
	\]
	
	\textbf{Scenario 1 (Watch-Time Priority):} $\alpha=1.0, \beta=2.0, \gamma=1.5, \delta=1.0$. \cite{covington2016deep}
	$U_{\text{Collab}} = 16.5$. $U_{\text{Beef}} = 12.0$. Equilibrium: Collaboration. \cite{hron2023modeling, zheng2023agency}
	
	\textbf{Scenario 2 (High Sponsor Penalty):} $\delta=2.5$. \cite{zheng2023agency}
	$U_{\text{Collab}} = 16.5$. $U_{\text{Beef}} = 7.5$. Equilibrium: Collaboration strongly preferred.
	
	\textbf{Scenario 3 (Click/Share Priority):} $\alpha=2.5, \beta=0.5, \gamma=2.0, \delta=1.0$. \cite{bakshy2015}
	$U_{\text{Collab}} = 13.5$. $U_{\text{Beef}} = 18.5$. Equilibrium: Beefing. \cite{verwiebe2024mercurial}
	
	\section{Policy Implications}
	The algorithm's weights $(\alpha, \beta, \gamma)$ dictate creator strategy \cite{covington2016deep}.
	\begin{itemize}
		\item \textbf{Incentivize Watch Time:} High $\beta$ promotes collaborative, long-form content \cite{covington2016deep}.
		\item \textbf{Incentivize Clicks:} High $\alpha, \gamma$ provokes conflict \cite{verwiebe2024mercurial}.
		\item \textbf{Sponsor Leverage:} Increasing $\delta$ algorithmically penalizes drama and prevents toxicity \cite{zheng2023agency}.
		\item \textbf{Partial Transparency:} Disclosing general priorities without exact formulas prevents manipulation of the algorithm \cite{bakshy2015}.
	\end{itemize}
	
	\section{Case Studies}
	\subsection{Collaboration (MrBeast)}
	MrBeast uses collaboration in order to minimize $\text{DramaRisk}_s$ while also maximizing $\beta$ which represents watch time, and $\gamma$ which represents shares. This allows him to gain high creator utility and draw zero sponsor penalty, generating a stable, cooperative Stackelberg equilibrium \cite{hron2023modeling, elberse2023mrbeast, millerhogg2023mrbeast}.
	
	\subsection{Conflict (Logan Paul vs. KSI)}
	Public fights or feuds maximize  $\alpha$ (clicks) and $\gamma$ (shares) however the trade off is that they incur severe sponsor penalties. Creators adopt this strategy when algorithmic rewards exceed sponsor risks. Platforms can curb this behavior by artificially inflating $\delta$ \cite{verwiebe2024mercurial, zheng2023agency, timeloganpaul2018}. 	One instance of delta firing was after Logan Paul's infamous 2018 "Suicide Forest" video, in which YouTube removed him from Google Preferred which was its top-tier ad-sales tier, and pulled him from a YouTube Red original movie, and dropped him from season 4 of "Foursome." Thus proving a direct platform-level Drama Risk penalty that is completely different from regular, organic audience backlash. It should be noted that his subscriber count and view numbers kept climbing throughout the controversy, which is consistent with this paper' model: specifically the algorithm's engagement reward (clicks, shares) and the sponsor's drama penalty are separate, sometimes opposing, forces acting on the same strategy choice.

	\section{Limitations}
	\begin{itemize}
		\item \textbf{Binary Strategy Space:} Forcing choices into absolute "collaboration" or "beefing" ignores continuous hybrid strategies \cite{krishna2009game}.
		\item \textbf{Homogeneous Creators:} Assuming identical creator utility ignores the impact of audience size and pre-existing reputation \cite{gibbons1992primer}.
		\item \textbf{Static Environment (partially addressed):} The baseline model is a one-shot interaction. Section~3 introduces a dynamic extension with period-indexed viewer fatigue and compounding sponsor penalties, but stops short of a full repeated game with evolving algorithmic weights \cite{fudenberg1991game}.
		\item \textbf{Linear Utility:} Sponsor penalties and engagement returns are likely non-linear \cite{mas1995microeconomic}.
		\item \textbf{Passive Agents:} Viewers and sponsors lack active strategies \cite{camerer2004advances}.
		\item \textbf{Single-Platform:} Creators cross-optimize across multiple platforms simultaneously \cite{tufekci2015}.
	\end{itemize}
	
	\section{Conclusion}
	Social media ecosystems function as Stackelberg games. The algorithm (leader) sets engagement parameters; creators (followers) respond to maximize utility. We demonstrated that assigning high weight to clicks and shares incentivizes conflict, while prioritizing watch time and sponsor penalties fosters collaboration. A toxic platform environment is not inevitable; it is a mathematical consequence of poorly tuned incentives. 
	
	\section*{Acknowledgments}
	I would like to acknowledge two resources that were instrumental in helping me build the foundation for this paper.
	
First, I am grateful to Heinrich von Stackelberg's foundational work, Marktform und Gleichgewicht (1934). It was only through understanding and deconstructing bit by bit the original Stackelberg model, that I was able to fully comprehend the core leader-follower dynamic that I ultimately adapted and used for the platform-creator interaction within this paper and model. 

	Secondly, I would like to acknowledge the Yale University Open Course ''Game Theory'' (Econ 159), which was taught by Professor Ben Polak, which I was able to learn from on YouTube. It was vital as it allowed me to attain the knowledge and vocabulary needed to understand utility functions, best responses, and equilibrium ultimately leading me to being able to make my own model.

Finally, I would like to acknowledge the online mathematics learning resources that crucially supported my learning and ultimately my work. I was self-studying AP Calculus AB through Khan Academy before and while developing this model. In order to get a better handle on the multi-variable optimization which were key to my utility functions, I watched select lectures from MIT OpenCourseWare's Multivariable Calculus course. These resources helped me extend my understanding from single-variable functions to optimizing functions of several variables like $\alpha$, $\beta$, and $\gamma$, and clarified the notation and mechanics behind the arg max operator used and found throughout this work.

	\bibliographystyle{plain}
	\bibliography{references}
\end{document}